# THE SEARCH FOR FRACTIONAL CHARGE ELEMENTARY PARTICLES AND VERY MASSIVE PARTICLES IN BULK MATTER[*]

Martin L. Perl, Valerie Halyo, Peter C. Kim, Eric R. Lee,
Irwin T. Lee, and Dinesh Loomba

*Stanford Linear Accelerator Center*
*Stanford University, Stanford, CA 94309*

## ABSTRACT

We describe our ongoing work on, and future plans for, searches in bulk matter for fractional charge elementary particles and very massive elementary particles. Our primary interest is in searching for such particles that may have been produced in the early universe and may be found in the more primeval matter available in the solar system: meteorites, material from the moon's surface, and certain types of ancient terrestrial rocks. In the future we are interested in examining material brought back by sample return probes from asteroids. We will describe our experimental methods that are based on new modifications of the Millikan liquid drop technique and modern technology: micromachining, CCD cameras, and desktop computers. Extensions of our experimental methods and technology allow searches for very massive charged particles in primeval matter; particles with masses greater than $10^{13}$ GeV. In the first such searches carried out on earth there will be uncertainties in the mass search range. Therefore we will also discuss the advantages of eventually carrying out such searches directly on an asteroid.

Paper presented at the Workshop on Cosmic Genesis and Fundamental Physics,
Sonoma State University, October 28 – 30, 1999.

[*] Work supported by Department of Energy contract DE-AC03-76SF00515.

# I. THE SEARCH FOR FRACTIONAL CHARGE ELEMENTARY PARTICLES PRODUCED IN THE EARLY UNIVERSE

## A. Search Motivation

Our ongoing research interest is in searching in bulk matter for fractional charge elementary particles produced in the early universe – particles whose properties preclude discovery using accelerators or other traditional search methods of high energy physics. When a star forms, these particles will be pulled into the star as either free particles or attached to light nuclei. Upon the disintegration of the star, these particles, either free or attached to nuclei, would be expelled into space. Our hope is that these particles would eventually be swept up in the formation of the solar system.

There are two benefits in searching for fractional charge particles through this path. First, the early universe production processes for such particles are very general, especially with respect to the mass range. Second, even an upper limit on the occurrence of fractional charge particles tells us about the constituents of the solar system in particular and the universe in general.

## B. Review of Search Methods for Fractional Charge Particles

Since Millikan's determination ninety years ago of the electron's charge using falling liquid drops in an electric field [1], there have been numerous searches for free particles with fractional electric charge. Searches have been carried out using accelerators [2,3,4], cosmic rays [2,3], indirect bulk matter searches [2,3,5], and direct bulk matter searches [6–17]. There are *no confirmed* discoveries of free particles with fractional charge. For example, La Rue *et al.*[7] claim to have found fractional charge in niobium, but that claim was disputed by the more extensive search in niobium by Smith *et al.* [6,10]. Of course, conventional particle theory holds that single quarks have fractional charge but also holds that free, single quarks to not exist.

As discussed in Ref. 15, accelerator searches and indirect bulk matter searches based on accelerators have limited significance because the range of accessible particle masses is limited by the accelerator's maximum energy; furthermore a production mechanism and cross section must be assumed. Cosmic ray searches are similarly limited in significance [15].

If one believes, as we do, in the possibility of stable [18] fractional charge particles having been produced in the early universe, then searches in bulk matter are always significant. In a bulk matter search one examines a sample of total mass $m$ of some type of matter, and either finds fractional charge particles, or gives an upper limit on the number of fractional charge particles per nucleon in that type of matter. Of course, finding fractional charge particles has great significance. Establishing an upper limit is also significant, particularly so, if two further conditions are met.

First significant consideration – the limit is more significant if the search does *not* include a concentration process using chemical, evaporative, or other method. As discussed by Lackner and Zweig [19,20], the chemical properties of atoms or molecules containing fractional charge particles are altered in complicated ways. Therefore it may be difficult to take an upper limit found after concentration and to deduce from it an upper limit for the original bulk material. We use the term indirect for bulk matter searches involving a concentration process, the remainder of this paper is devoted to direct matter searches, that is searches where all of the mass $m$ is examined.

Second significant consideration – most bulk matter searches, Table I, have examined substances that have gone through chemical transformation through manufacturing, refining, or geochemical processing. There is usually uncertainty as to how well a fractional charge particle attached to an atom or molecule will be carried through such processes [15,19]. Therefore, there may be diminished significance for upper limits in refined or manufactured substances such as iron, niobium, or silicone oil.

Returning to Table 1, we see that individual published bulk matter searches have examined less than 5 mg of any one material. Note that 1 mg contains about $6 \times 10^{20}$ nucleons. *All* searches had negative results, except LaRue *et al*. As discussed in Sec. 2B, we [17] have concluded, and are publishing, a search in about 17 mg of silicone oil.

As listed in Table 1, there are two different methods for bulk matter searches. In the levitometer method [2,3,7-11,15] a ball [21] of the substance to be examined, about 0.03-0.1 mg in mass, is levitated using ferromagnetism or superconductivity. The ball's charge is measured using an oscillating electric field.  In each of the levitometer searches listed in Table 1, ten to a hundred different balls are examined.

Table I  Published results for bulk matter searches for fractional charge particles.  All searches except that by LaRue *et al.*, [7] reported no evidence for fractional charge particles in these examples.  Note that 1 mg contains about $6\times10^{20}$ nucleons.

| Method | Experiment | Material | Sample Mass (mg) |
|---|---|---|---|
| Superconducting levitometer | LaRue *et al.* [7] | niobium | 1.1 |
| Ferromagnetic levitometer | Marinelli *et al.* [8] | iron | 3.7 |
|  | Smith *et al.* [10] | niobium | 4.9 |
|  | Jones *et al.* [11] | meteorite | 2.8 |
| liquid drop | Joyce *et al.* [13] | sea water | 0.05 |
|  | Savage *et al.* [12] | native mercury | 2.0 |
|  | Mar *et al.* [16] | silicone oil | 1.1 |

In our preferred method, liquid drops are used of much smaller masses than the levitometer balls.  The charge of the drop is determined by measuring the terminal velocity of the drop moving through air under the influence of an electric field, Sec. 2.  The use of a very large number of drops, $4.1\times10^7$ in our last measurement [17], provides natural self-calibration of the charge measurement, enables detailed study of measurement errors, and allows a variety of substances to be studied [22].

## C.   Fractional Charge Particle Searches in Primitive Materials

Our past searches for fractional charge particles [16,17] have been confined to silicone oil while we developed the experimental methods described in this paper, Sec. 2.  However, our primary interest is in searching for fractional charge particles in meteorites derived from asteroids.  These meteorites are the most primitive solar system materials available on earth; therefore, the search for fractional charge particles in these materials is as close as we can come at present to a general search in the solar systems and universe,

Sec. 1.A. The only past search of meteoritic material is that of Jones *et al.* [11], and no evidence for fractional charge particles was found.

We also have a strong interest in certain terrestrial materials that may more readily accumulate atoms containing, or bound to, fractional charge particles. This interest is based on the work of Lackner and Zweig [19,20].

## II. TRADITIONAL AND NEW LIQUID DROP SEARCH METHODS

### A. Traditional Millikan Liquid Drop Method

Our first two searches [16,17], both in silicone oil, used what we call, the traditional Millikan liquid drop method, Fig. 1. Drops with diameters in the range of 6 – 12 μm are produced using a piezoelectric drop generator of our own design, Fig. 2 [23]. The exact diameter depends on how we set the parameters of the generator: aperture size, aperture shape, pulse shape, and pulse frequency. Once these generator parameters are set, the drop diameter is maintained to about 0.1%.

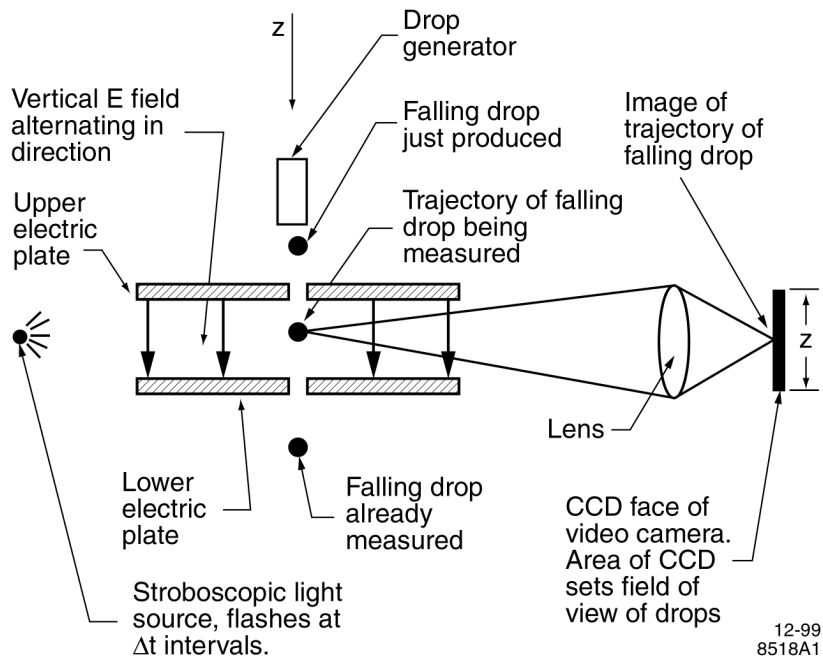

Fig. 1. Schematic vertical view of the traditional Millikan liquid drop apparatus.

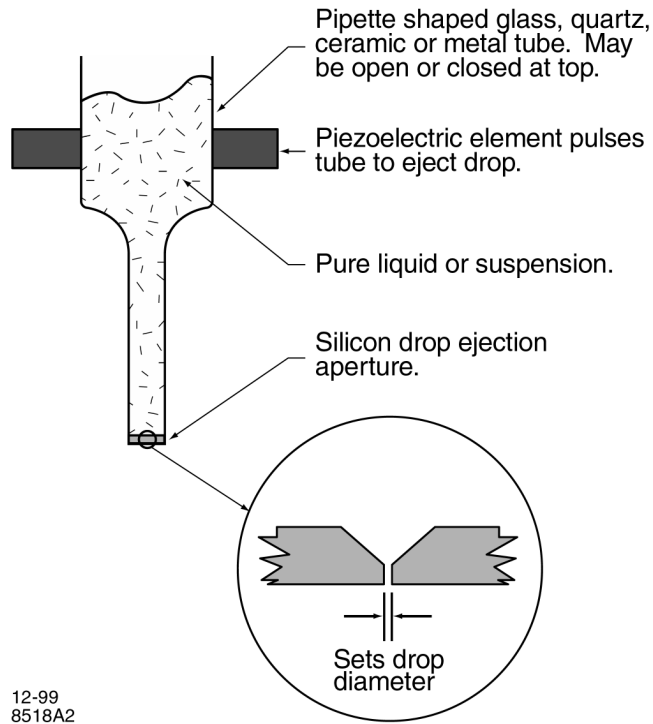

Fig. 2. Explanatory view of the piezoelectric drop generator. The diameter of the hole in the drop ejection aperture plate sets the diameter of the drop within range of about ±15%. The exact diameter depends upon aperture shape, pulse shape, and pulse frequency.

The drops fall through air, the forces on the drops are: that due to gravity, that due to the resistance of the air, and that due to a vertical electric field. The electric field points alternately up and down. When the field points down the drop terminal velocity is $v_{down}$, when the field points up the terminal velocity is $v_{up}$. The terminal velocities are determined by measuring the vertical position, $z$, of the drop at time intervals of $\Delta t$. In our use of the traditional Millikan method we make of the order of ten to twenty successive determinations of the vertical position, $z_1, z_2, z_3…$; this increases the precision of our knowledge of the terminal velocities and checks that the drop charge has not changed during the measurement period.

The terminal velocities are given by the equations:

$$\frac{4}{3}\pi r^3 \rho g + QE = 6\pi \eta r v_{down}$$
$$\frac{4}{3}\pi r^3 \rho g - QE = 6\pi \eta r v_{up}. \tag{1}$$

Here, $r$ is the drop radius, $\rho$ is the mass density of the drop less the mass density of air, $Q$ is its charge, $E$ is the magnitude of the electric field (taken to be the same for both directions), and $\eta$ is the viscosity of air. We calculate the charge $Q$ and radius $r$ of each drop from the measured values of $v_{down}$ and $v_{up}$, using Eq.1.

Fig. 3a, based on part of the data acquired in Ref. 17, shows the typical charge distribution that we find in *silicone oil*. Here the charge is defined in units of the electron charge, $e$, namely $q = Q/e$. This data set contains $2.5 \times 10^7$ drops of about 10 μm diameter, the total mass of this sample is about 10 mg. We see that in this data set the measurements all fall in narrow peaks about $q = 0, \pm 1, \pm 2, \ldots$

These peaks are more easily examined in Fig. 3b by defining a residual charge in terms of the integer peak centers, $q_c = q - N_{nearest}$, where $N_{nearest}$ is the nearest signed integer. A Gaussean distribution fits the combined peaks with a standard deviation of $\sigma_Q = 0.020$ e, that is, the drop charged is being measured with a $\sigma_Q$ of about 1/50 of an electron charge. About half of $\sigma_Q$ comes from the Brownian motion of a drop in the air, the other half comes from measurement errors.

In Ref. 17 we present and discuss the full data, consisting of three data sets using drops of three different diameters, and totaling 17 mg. All but one of the $4.1 \times 10^7$ drops have charges that fall close to integer values within the $\sigma_Q$'s quoted in the previous paragraph. The possible significance of the one drop with anomalous charge is discussed in Ref. 17. Our conclusion is that there is no evidence for fractional charge particles in this 17 mg sample of silicone oil [17].

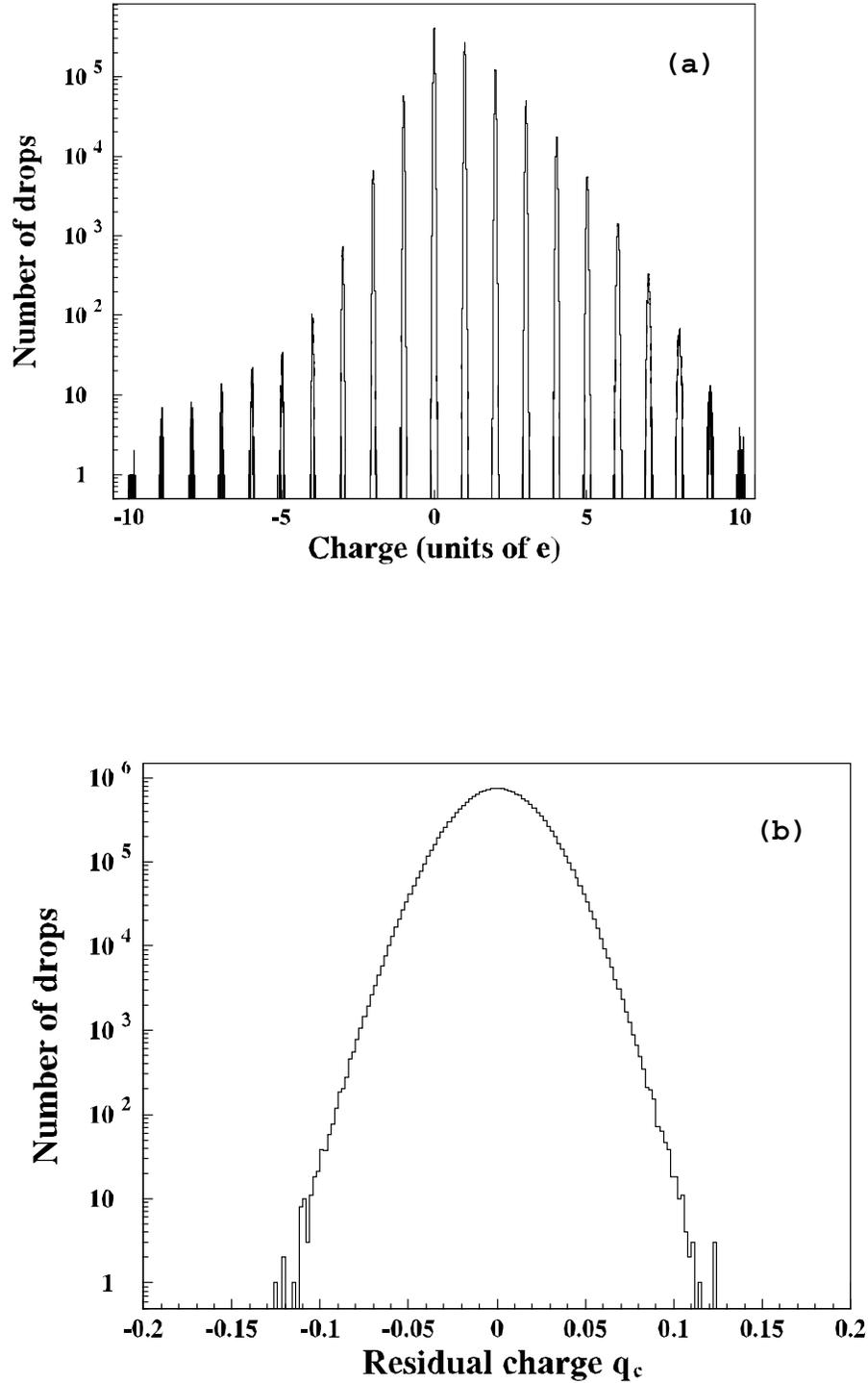

Fig. 3. (a) The charge distribution in terms of $q = Q/e$ for $2.5 \times 10^7$ drops of silicone oil of diameter about 10 μm. (b) The residual charge distribution with respect to the peak centers, $q_c = q - N_{\text{nearest}}$, showing the superimposed peaks.

**B.     Searching for Fractional Charge Particles in Mineral Suspensions and the Large Drop Problem with the Traditional Millikan Method**

In preparing suspensions of meteoritic or other mineral materials for the liquid drop search method we do *not* chemically process the material in any way.  For example, we do not dissolve the minerals in acid and then attempt to form a fine precipitate.  The reason for our ban on chemical processing is that we have no way of knowing whether or not we might lose a fractional charge particle during the processing.  There are several loss mechanisms in this type of process:  the charge particle might be collected by the walls of the processing vessels; the charged particle, with or without its associated atoms, might not go into solution; or the charged particle, with or without its associated atoms, might not precipitate.

Our first step in pulverizing mineral samples is to use a ball mill or a mortar and pestle.  A rule of thumb in pulverization technology is that such methods can only reduce many minerals to small pieces whose minimum size is of the order of ten or several tens of μms.  There are two reasons, first minerals break at defects such as grain boundaries, and the smaller the piece, the less chance it contains defects.  Second, as the mineral pieces get smaller they protect each other against further milling.

Our second step in pulverization is to use a jet pulverizer, a device in which high-speed air jets entrain the powder from the first step; the mineral pieces being given relatively high kinetic energies.  These pieces then collide with each other further reducing their sizes.  In minerals with which we work, this produces a powder with sizes ranging from about 0.1 μm to about 5 μm.  It is our experience, and those of others, that further passes through the jet pulverizer *do not* substantially change the size distribution.

We find that when we generate 10μm diameter drops containing a suspension of powdered mineral with powder sizes in the 0.1 μm to 5 μm range, the drop generator, Fig. 2, works poorly.  Sometimes it stops working completely; sometimes the mineral powder remains in the ejector tip with the generated drops containing little or no mineral powder.

Of course we could use just the fine end of the 0.1 μm to 5 μm powder to make a suspension, using say just 0.1 μm to 1μm powder.  But this would lead to a biased search of the mineral for fractional charge particles.  Recall that the minerals we use, such as meteorites, are a complicated mixture of many different compounds such as silicates and oxides of different elements. As an extreme example, suppose that a particular compound in the mineral is the best vehicle for entraining fractional particles, and suppose that compound is the hardest to pulverize.

Our solution to this experimental dilemma is obvious, we use larger drops, about 20 µm or larger in diameter. Then, with mineral powders in the size range of 0.1 µm to 5 µm we can get consistent generation of drops containing suspensions of these minerals. But there is a practical problem in using drops larger than about 15 µm in diameter in the traditional Millikan liquid drop method [22]. The average terminal drop velocity from Eq. 1 is

$$v_{ave} = \frac{2r^2 \rho g}{9\eta}. \tag{2}$$

Referring to Fig. 1, the total time available for observing a falling drop is

$$t_{total} = \frac{Z}{M v_{ave}} \propto \frac{Z}{M r^2}. \tag{3}$$

Here $Z$ is the total distance the drop falls while being viewed by the CCD camera and $M$ is the magnification from the drop plane to the plane of the CCD face. Therefore, as $r$ increases there is a rapid decrease in the number of measurements that can be made of $v_{down}$ and $v_{up}$; hence $\sigma_q$ increases.

In principle, the problem of using larger drops in the traditional Millikan method apparatus can be solved by using an optical system and CCD camera with a much larger number of pixels along the $Z$ direction and by increasing the distance between the electric plates in Fig. 2. In practice this solution has many problems, for example, such a fast affordable CCD camera does not exist. Furthermore, increasing the distance between the plates while maintaining $E$ constant requires switching a larger voltage with concomitant difficulties.

## C. The Horizontal Electric Field Liquid Drop Method

Our solution to the large drop problem has two parts. First, as shown in Fig. 4, we decouple the effect of the electric field on the drop motion from the effect of the gravitational field by using a horizontal electric field.

In this configuration the motion in the vertical direction, $z$, depends on $r$ and $\rho$, and the motion in the horizontal direction, $x$, depends on $r$ and $q$:

$$v_x = QE/6\pi\eta r, \tag{4a}$$

$$v_z = 2r^2\rho g / 9\eta. \tag{4b}$$

Thus the charge is computed solely by measurement of $v_x$ using

$$Q = 6\pi\eta r v_x / E, \tag{5}$$

$r$ being obtained from the *image of the drop* on the CCD.

In contrast to the traditional liquid drop method where $v_z$ is needed to determine $r$ as in Eqs. 1 and 2; $v_z$ is now *not* of interest. This lends to the second part of the solution of the large drop problem. We use laminar, upward directed airflow against the direction of the gravitational force. Let $v_{term.drop}$ be the terminal velocity of the drop relative to the moving air and let $v_{air}$ be the upward velocity of the air. Then the downward velocity of the drop relative to the apparatus is

$$v_{z\,drop} = v_{term\,drop} - v_{air} \tag{6}$$

Consider the following example. A 20 $\mu$m diameter drop with $\rho = 1$ g/cm$^3$ has $v_{term\,drop} = 12$ mm/s. It is easy to produce a laminar airflow of $v_{air} = 11$ mm/s. Hence $v_{z\,drop} = 1$ mm/s, producing plenty of observation time for measuring $v_x$ and $r$, and hence $q$. Indeed the only use of the gravitational field is to continually pull drops out of the space where $v_x$ is measured.

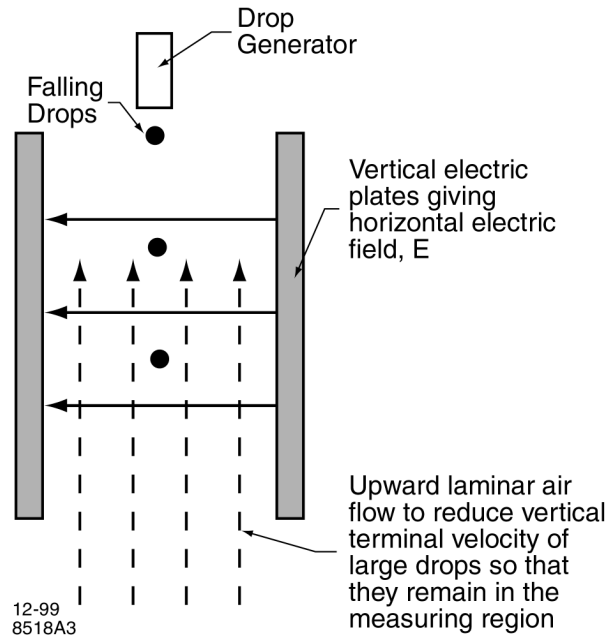

Fig. 4. Schematic illustration of the use of a horizontal electric field and the use of upper, laminar airflow. The purpose of the airflow is to reduce drastically the downward vertical drop velocity, $v_z$ drop relative to the apparatus. The view is looking along the axis of the optical system.

At present we are putting into operation this fractional charge research method - a horizontal electric field and upward airflow. We are using drops with diameters in the 20 to 30 μm range. Details are given in Ref. 22 as well as some alternatives to the use of upward airflow.

**D.     Our Near Term Goals for Searches for Fractional Charge Particles**

We have the following goals for our next searches for fractional charge particles. These searches will use our new horizontal electric field, liquid drop method.

1.   We will repeat our search in silicone oil with a larger sample.
2.   Our major next step is to study unrefined materials, particularly meteorites. We have samples from the Allende meteorite.
3.   We will also study fluorapatite, a mineral that collects fluorine-like elements.

4. Sample sizes will be increased from tens of milligrams to perhaps a hundred milligrams.

**E. Twenty Five Year Goals for Searches for Fractional Charge Particles**

One of the motivations for the Workshop on Cosmic Genesis and Fundamental Physics was to discuss twenty five year goals of new research directions. We see two grand goals for fractional charge particles.

1. We would like to increase the sample size in a search from tens of milligrams to tens of grams or even a kilogram. We do not know what will be required in improvements on our present liquid drop method or in the invention of a new bulk search method.
2. We would like to bring large samples from asteroids.

**III. BULK MATTER SEARCHES FOR VERY MASSIVE PARTICLES PRODUCED IN THE EARLY UNIVERSE**

**A. Search Motivation**

The upper mass limit on direct searches for massive particles at the 14 TeV Large Hadron Collider (LHC) will be about one fifth the total energy, namely $3 \times 10^3$ GeV/c$^2$. Indirect particle searches based on resonant or interference effects in production processes will reach to larger masses by one to two powers of ten, about $10^5$ GeV/c$^2$. Proposals for a very large hadron collider [24] project a total energy of 100 to 200 TeV, extending the mass reach to $10^6$ GeV/c$^2$.

Is there any way to search for yet more massive particles? Surprisingly, one can search in bulk matter for a class of such particles using a falling drop method [25]. The criteria for particles in this class are:

1. mass in the range of $10^{13}$ to $10^{17}$ GeV/c$^2$. Of course these particles would have to be present in the solar system through the same path outlined for fractional charge particles in Sec. I.A.,
2. produced in the early universe,
3. stable, and

4.  charged or bound by the strong interaction to a stable charged particle.

**B.  Liquid Drop Search Method**

This method depends upon some mass relationships. The mass of a 6 μm diameter drop with a typical mineral suspension of density 1.4 grams/cm$^3$ is

$m_{drop} \approx 1.6 \times 10^{-10}$ grams.

Since

1 GeV/c$^2$ = $1.8 \times 10^{-24}$ grams,

$m_{drop} \approx 10^{14}$ GeV/c$^2$.

Thus our smaller drops have a mass equal to or less than particles that might exist in the interesting mass range of $10^{14}$ GeV/c$^2$ and above.

Consider an apparatus that measures the terminal velocity of drops falling in air. It would be simpler than that in Fig. 1 because there is no electric field. A drop of mass m has terminal velocity $v(m)$:

$v(m) = mg/6\pi\eta r$.

Where $g$ is the acceleration of gravity, $\eta$ is the viscosity of air, and $r$ is the drop radius. Suppose a drop also contains an elementary particle of mass M, then the terminal velocity is

$v(m+M) = (m+M) g/6\pi\eta r$.

Figure 5, an illustrative plot of number of drops $N$ versus $v$, shows what we hope to see: a very large peak in $dN/dv$ at $v(m)$ and a relatively very small peak at $v(m+M)$. Our ability to detect the $v(m+M)$ peak depends on the abundance of the massive particle and on the width and tails of the $v(m)$ peak. As a first estimate we believe we can separate the $v(m)$ and $v(m+M)$ peaks if M ≥ m.

There is a potential problem that we must investigate by experiment. Drops consisting of a suspension of pulverized meteoritic or other material will not have uniform density. This non-uniform density may produce extended tails on the $v(m)$ peak; the tail at larger $v$ values is a serious problem if it extends to $v(m+M)$.

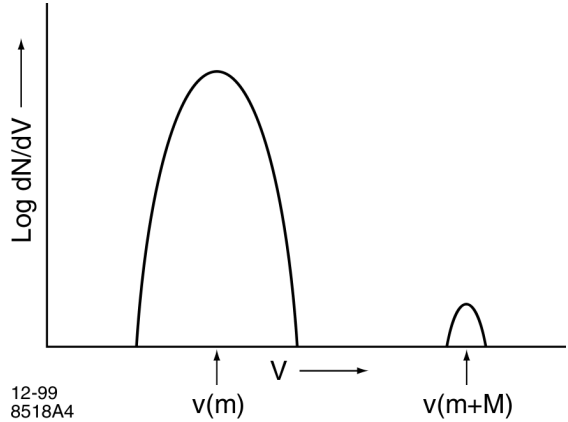

Fig. 5. Illustrative plot of number of drops, $dN/dv$, versus the terminal velocity $v$, shows what we hope to see if some drops contain a massive particle of mass $M$: a very large peak at $v(m)$ and a relatively very small peak at $v(m+M)$. The mass of the drop is $m$.

## C. Lower and Upper Mass Limits in Search for Very Massive Particles

The lower mass limit of the search method is determined by the rough requirement $M \geq m$ and by the minimum size drops we can use in a practical experiment. We can probably reliably produce drops with 4 µm diameter, giving a lower limit on $M$ of about $10^{13}$ GeV/c$^2$. We do not see how to extend this method to yet smaller drops: it may be difficult to make such drops reliably, it will be difficult to get reliable measurements of the drop radius $r$, and it will be difficult to search through large amounts of material.

As discussed in Ref. 25, the mass limit comes from the necessity in this search method of the massive particle remaining bound in ordinary matter while in Earth's gravitational field. There must exist a binding force $F_b$ between the particle and the drop's ordinary matter so that $F_b$ is larger than the gravitational force on the particle, $Mg$. The straightforward binding mechanism is electric charge. We suppose the particle is charged or is bound by the strong force to a charged particle. To estimate $F_b$, we suppose

(a) the massive particle has an electric charge $e$, where $e$ is the electron charge, (b) the binding energy to the ordinary matter is about 1 eV, and (c) $F_b$ extends over about $10^{-10}$m. Then $F_b$ is about $1.6 \times 10^{-9}$ nt, and $M$ must be less than $F_b/g \approx 1.6 \times 10^{-10}$ kg $\approx 10^{17}$ GeV/c$^2$.

Hence this proposed search for massive stable particles with electric charge could extend from $10^{13}$ to $10^{17}$ GeV/c$^2$. There is certainly some optimism in the calculation of these limits. The lower limit might not be quite so low if it proves to be difficult to use drops of less than 6 μm diameter. The upper limit might not be quite so high if the particle has fractional electric charge or we have been too generous in estimating the strength of $F_b$.

**D.     Our Near Term Goals for Searches for Very Massive Particles**

In the course of developing this search method we will use a terrestrial mineral sample. But the geological history of the earth is complicated and particles in the $10^{13}$ to $10^{17}$ GeV/c$^2$ mass ranges may have long since moved to the earth's center.

The best materials for very massive particle searches are meteorites from asteroids and it is here that we shall put our first serious effort. Unfortunately, there is a problem with the upper mass limit when searching meteorites. As pointed out by Jean and Longo [26], when meteorites enter the atmosphere they slow down, the deceleration force may be 100g to 1000g. Therefore the more massive particles will *not* stay in the meteorite.

**E.     Twenty Five Year Goals for Searches for Very Massive Particles**

There are two solutions to the meteorite deceleration problem. One solution is to bring back asteroid samples by a small acceleration and small deceleration orbit, perhaps keeping the acceleration or deceleration to less than 10g.

The other solution, grand and exciting, is to send the massive particle search apparatus to an asteroid, carrying out the search on the asteroid. There are three great advantages.

1.     There are no particle loss problems from acceleration or deceleration.
2.     Since $g_{asteroid} \ll g_{earth}$ the upper mass limit for searches is increased.
3.     Since $g_{asteriod}$ is relatively small, very massive particles may lie on the surface.

We don't know if we can turn this dream into a reality, the technical problems are hard, but we don't know of any other way to search for very massive particles.

ACKNOWLEDGEMENTS

This work was supported by the Department of Energy contract DE-AC03-76SF00515.